# Molecular fingerprint-region spectroscopy from 5–12 µm using an orientation-patterned gallium phosphide optical parametric oscillator

Luke Maidment,[1,*] Peter G. Schunemann,[2] and Derryck T. Reid[1]

[1] *Scottish Universities Physics Alliance (SUPA), Institute of Photonics and Quantum Sciences, School of Engineering and Physical Sciences, Heriot–Watt University, Edinburgh EH14 4AS, UK*
[2] *BAE Systems, Inc., MER15-1813, P.O. Box 868, Nashua, NH, USA 03061-0868*
*Corresponding author: lm345@hw.ac.uk*

**We report a femtosecond optical parametric oscillator (OPO) based on the new semiconductor gain material orientation patterned gallium phosphide (OP-GaP), which enables the production of high-repetition-rate femtosecond pulses spanning 5–12 µm with average powers in the few to tens of milliwatts range. This is the first example of a broadband OPO operating across the molecular fingerprint region, and we demonstrate its potential by conducting broadband Fourier-transform spectroscopy using water vapor and a polystyrene reference standard. © 2015 Optical Society of America**

Molecular absorption spectra in the 3–20-µm infrared wavelength range contain intense and distinctive patterns of features, providing a unique 'molecular fingerprint', which historically has been exploited to infer the composition of unknown chemicals by using Fourier-transform spectrometers illuminated with black-body radiation. Substituting thermal emitters with broadband coherent sources promises exciting new possibilities for molecular spectroscopy: temporal coherence is critical for mid-infrared frequency combs [1,2] and their applications in dual-comb spectroscopy [3–5], while spatial coherence enables diffraction-limited focusing in applications like near-field micro-spectroscopy [6] and free-space propagation for stand-off chemical detection [7]. The wide instantaneous bandwidth of mid-infrared femtosecond oscillators is particularly appealing, since it permits powerful Fourier-transform techniques to be adopted, which transfer the burden of precision wavelength calibration from the source to the detection system [8] while also providing excellent signal:noise characteristics and wavelength-independent spectral resolution [9].

With the exception of recent results using supercontinuum generation in chalcogenide fiber [10], technical routes to broadband coherent sources operating across the fingerprint region have been limited to synchrotrons [11,12], difference-frequency mixing between two near-infrared lasers [13] or different spectral components of broadband near-infrared pulses [14], and cascaded OPOs [15]. These approaches each face different limitations associated with their complexity, low repetition rate, size or poor efficiency, limiting their potential for widespread adoption. OPOs offer a compelling route to accessing the fingerprint region, and the most promising demonstrations to date have employed the quasi-phasematched (QPM) semiconductor gain material orientation-patterned gallium arsenide (OP-GaAs) [16], which has been shown to address 4–14 µm with nanosecond pulses [15]. Two-photon absorption prevents OP-GaAs from being pumped at wavelengths below 1.7 µm, and consequently its implementation in a femtosecond oscillator has been limited by the availability of suitable pump lasers to the production of idler wavelengths shorter than the fingerprint band [17].

Orientation-patterned gallium phosphide (OP-GaP) [18], a recently developed material and the first new orientation-patterned semiconductor to produce meaningful output powers in nearly 15 years [19], now presents an exciting solution to the problems outlined earlier. Unlike OP-GaAs, its band edge lies in the visible region, so two-photon absorption is negligible above 1 µm, while its transparency extends to 12 µm, providing excellent coverage of the fingerprint region. These characteristics and its substantial nonlinearity of 70.6 pm V$^{-1}$ [20] allow it to be pumped directly with powerful 1-µm lasers, and the domain patterning period—set in the growth procedure by the molecular beam epitaxy (MBE) template—provides the required tuning parameter needed to cover the material's full transparency range. Fig. 1 presents the phasematching efficiency map for a 1-mm-long OP-GaP crystal, along with the transmission spectrum of OP-GaP. The wavevector mismatch $\Delta k = k_{pump} - k_{signal} - k_{idler} - 2\pi/\Lambda$ was calculated over a matrix of grating periods (Λ) and signal / idler wavelengths and used to obtain the phasematching efficiency factor $\text{sinc}^2(\Delta k L/2)$, where L is the crystal length of 1 mm. This calculation was performed multiple times for pump wavelengths from 1020–1060 nm at 1 nm intervals and the results weighted by the pump laser spectral density to produce the final conversion-efficiency map. This calculation illustrates the suitability of using 1-mm crystals for the OPO. It can be seen that idler generation is possible throughout the OP-GaP transparency window by varying the quasi-phasematching period from 15–35 μm, values which are compatible with the capabilities of the hydride vapor phase epitaxy (HVPE) growth technique used to extend the thickness of the MBE-grown QPM layer to several hundred microns.

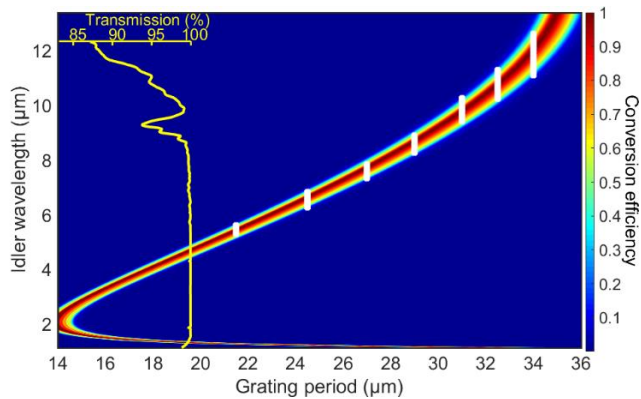

Fig. 1. Phasematching efficiency for a 1-mm-long OP-GaP crystal, calculated using the Sellmeier equation from [21]. The crystal transmission is shown on the yellow axis (inset). The white vertical bars depict the experimentally observed -10-dB idler bandwidth.

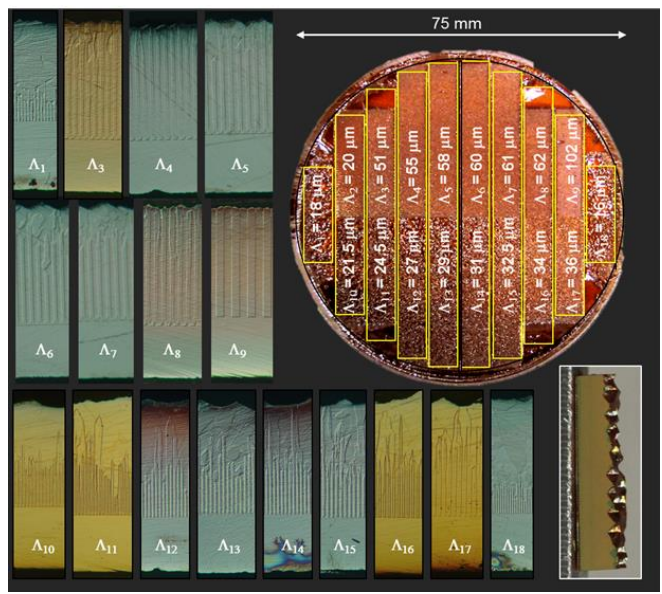

Fig. 2. The layout of the domain regions on the 75-mm OP-GaP wafer are shown together with cross sections revealing the domain-period growth. The inset on the bottom right shows a diced and anti-reflection coated crystal (grating vector normal to the page).

Based on this design, an OP-GaP wafer was prepared with eighteen distinct regions including patterning periods in the required range, as well as longer periods for evaluation purposes. The wafer was grown by the all-epitaxial processing technique originally developed at Stanford University [22,23] (and the University of Tokyo [24,25]) for the growth of OP-GaAs. This approach is based on the use of polar-on-nonpolar MBE to produce a III-V semiconductor layer whose orientation is inverted relative to that of the substrate. Beginning with a 3-inch-diameter undoped (100) GaP substrate (4-degree offcut toward <111B>), a 200-nm GaP buffer layer was grown prior to the deposition of the 5.6-nm non-polar silicon layer (performed in an auxiliary growth chamber), followed by the growth of a 5.6-nm AlGaP smoothing layer and finally a 200-nm GaP inverted layer free of anti-phase domains at the surface. The inverted layer was then photolithographically patterned with a multi-grating mask oriented with the domain stripes perpendicular to the [01-1] major flat, and alternating domains were reactive-ion etched in $BCl_3$ down to the starting substrate. After thoroughly removing the photoresist, the wafer was reloaded into the MBE chamber for 1000 nm of regrowth at a rate of 200 nm/hr.

To extend the thickness of the resulting 2-μm-thick QPM structure to millimeter-scale dimensions, the OP-GaP template was loaded into a horizontal, low-pressure HVPE reactor in which HCl gas flows over high-purity molten Ga to form GaCl vapor which exits from a showerhead directly above the rotating wafer and reacts with a flowing "sheet" of phosphine to form GaP according the reaction $GaCl + PH_3 \rightarrow GaP + H_2 + HCl$. At a growth temperature near 800 °C we achieved high-quality epitaxy (FWHM rocking curve peaks < 23 arcsec) at an average growth rate of 96 μm/hr (500 times faster than MBE) over an 8-hr run. The resulting wafer is shown in Fig. 2, and showed excellent vertical, parallel domain propagation even for the smallest periods (although these cease to propagate after a few hundred microns of growth for reasons that are not well understood). Useable crystals were obtained for each of the periods, with the growth typically extending over at least 150 μm.

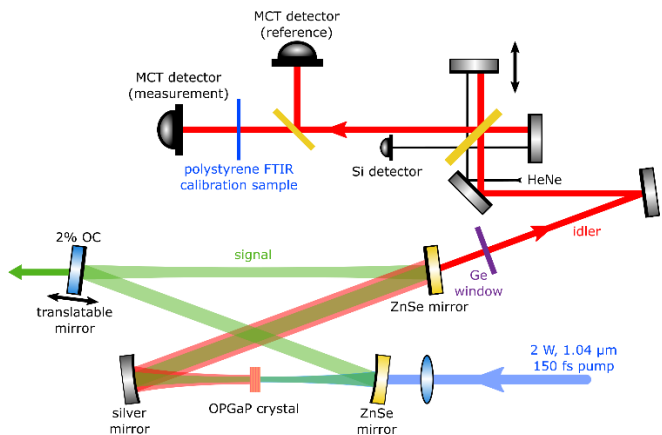

Fig. 3. Synchronously pumped OP-GaP OPO and Fourier-transform spectrometer.

OP-GaP crystals with lengths of 1 mm and several patterning periods were diced, polished, and anti-reflection (AR) coated for near- to mid-infrared wavelengths. We configured a synchronously pumped OP-GaP OPO in a 101.2-MHz resonator with high reflectivity from 1.15–1.35 μm (Fig. 3), pumped with 150-fs pulses from a 1040-nm femtosecond laser (Chromacity Spark). The coating of one spherical mirror was optimized for transmission at the pump wavelength of 1040 nm and for high reflectivity at the resonant signal wavelength in a range from 1.15–1.35 µm, while the other spherical mirror collimated the idler beam emerging from the OP-GaP crystal and was silver coated to provide high reflectivity for all idler wavelengths. This collimated idler beam was output-coupled from the cavity by transmission through a plane mirror coated with high transmission for the idler wavelengths (5–12 µm) and high reflectivity for the signal wavelengths (1.15–1.35 µm) on an infrared-transparent ZnSe substrate. The inclusion of a 2% signal output coupler, as well as the signal loss from the silver spherical mirror, allowed the idler power to be maximized by reducing back-conversion within the OP-GaP crystal. The OPO was brought into synchronism with the pump pulses by manual adjustment of the cavity length.

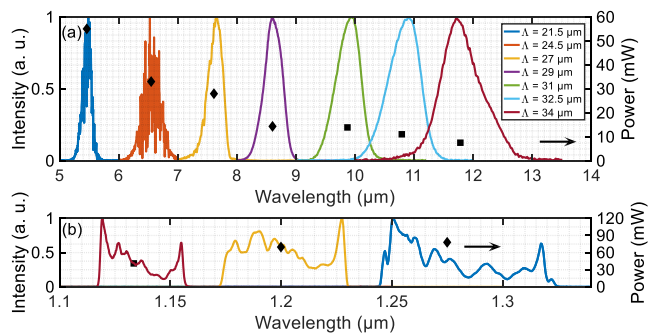

Fig. 1. (a) Idler spectra and (right axis) measured power. (b) Selected signal spectra and (right axis) measured power. Diamond markers represent measurements made using different cavity mirrors to the square markers.

A Michelson interferometer with one arm scanning at 2 Hz was used as a Fourier transform spectrometer. The speed of the scanning mirror was calibrated using a HeNe laser, providing an accurate wavelength scale for the measured mid-IR light (Connes' advantage of FTIR). Depending on the wavelength of the spectrum being measured, two different beamsplitters were used in the interferometer ($CaF_2$ AR coated for 2–8 µm, and ZnSe AR coated for 7–14 µm). The spectrometer resolution was limited by the scanning range of the interferometer to 1.2 cm$^{-1}$. A liquid nitrogen cooled mercury cadmium telluride (MCT) detector, sensitive from 3–12 µm, was used to record idler spectra from seven OP-GaP crystals with domain periods from 21.5–34.0 µm. Spectra centered from 5.4–11.8 µm and extending to 12.5 µm are presented in Fig. 4(a), together with average power measurements for each crystal. Atmospheric water absorption is evident in the two shortest wavelength spectra, while at longer wavelengths clean spectra are obtained since the 8–12-µm band corresponds to an atmospheric transmission window. The MCT detector sensitivity tails off at wavelengths > 11.8 µm, and this likely affects the shape of the Λ = 34 µm spectrum. A different mirror set with high reflectivity from 1.10 – 1.25 µm was used for the 31, 32.5 and 34-µm gratings. The maximum average power was 55 mW at 5.4 µm with 7.5 mW being recorded at 11.8 µm. Idler powers were recorded directly after an anti-reflection coated Ge window and do not account for losses from the silver mirror or the window. Fig. 4(b) presents signal spectra recorded for different domain periods, illustrating the modest tuning needed to provide extensive mid-infrared coverage. There is significant overlap of the spectra for adjacent gratings, so only three spectra are presented here for clarity. Data in Fig. 1 representing the -10-dB bandwidths of the idler spectra show that the Sellmeier equations used to calculate the OP-GaP phasematching [21] can accurately predict the observed tuning behavior, despite being measured at much longer wavelengths.

To demonstrate the practicality of this source for molecular fingerprint-region spectroscopy we implemented a dual-detector Fourier-transform spectrometer (Fig. 3) to acquire a reference spectrum synchronously with a measurement spectrum. Transmission spectra of a polystyrene reference sample were measured by simultaneously recording the spectrum transmitted through the polystyrene and a reference spectrum, then calculating their ratio. This approach prevents any drift in the idler spectrum from influencing the measurement. At shorter wavelengths we used atmospheric water vapor absorption as our spectroscopy reference, while in the atmospheric transmission window we used a polystyrene reference standard. Spectroscopy results are presented in Fig. 5, which includes the water vapor transmission spectra from the HITRAN molecular absorption database for comparison [26]. Excellent agreement is obtained across the 5–13-µm region, representing a significant portion of the fingerprint region containing many of the diagnostic stretching and bending frequencies of important molecular functional groups.

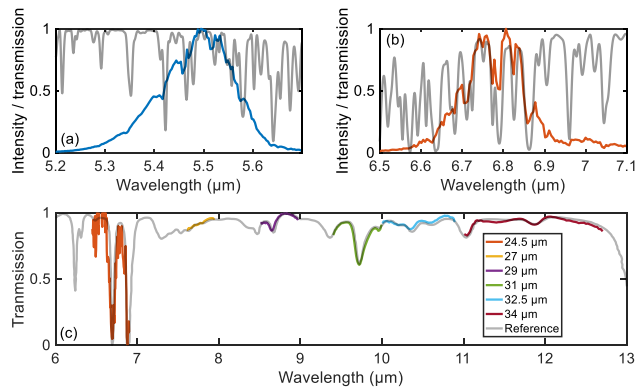

Fig. 2. Simulated water-vapor absorption spectra from the HITRAN database (grey) shown for comparison with idler spectra recorded from the OPO operated with OP-GaP grating periods of 21.5 µm (a) and 24.5 µm (b). (c) Transmission spectra of a polystyrene reference standard recorded from the OPO operated with OP-GaP grating periods of 24.5–34 µm and reference spectrum supplied by the sample manufacturer (grey).

The OP-GaP OPO is the first OPO to reach the molecular fingerprint region using direct 1-µm laser pumping, allowing it to exploit reliable and high power Yb laser technology, rather than less mature 2-µm pump sources. The average power is also extremely high: even above 10 µm, several milliwatts are produced, in comparison with 150 µW (full bandwidth) from a state-of-the-art mid-infrared fiber supercontinuum source [10] and 200 µW from a dual-laser difference-frequency generation source [13]. When spectral brightness or brilliance (power per unit optical frequency, accounting for beam area and divergence) is considered, the advantages of an OPO source are even more apparent over a alternative sources. The highest brilliance fingerprint-region source reported, to our knowledge, produces $4.9 \times 10^{19}$ ph s$^{-1}$ mm$^{-2}$ sr$^{-1}$ 0.1%BW$^{-1}$ at a center wavelength of 11.5 µm [14] (a 103-mW spectrum covering 6.7–18 µm). The method used in [14] to calculate the brilliance (also referred to as spectral brightness) was implemented for the OP–GaP OPO as follows. At a central wavelength of 11.8 μm an average output power of 7.5 mW was measured. The power contained within a bandwidth of 0.1% of the central wavelength was 0.10 mW, or $6.1 \times 10^{15}$ photons s$^{-1}$ 0.1%BW$^{-1}$. The signal beam waist in the OPO was 19 μm (from cavity stability matrix calculations) and we assumed a diffraction limited idler beam with the same waist size. The idler beam area ($1.13 \times 10^{-3}$ mm$^2$) and divergence (197 mrad) gave a solid angle of 0.12 steradians and so implied a brilliance of $4.1 \times 10^{19}$ ph s$^{-1}$ mm$^{-2}$ sr$^{-1}$ 0.1%BW$^{-1}$. The OP-GaP OPO does not cover such a broad range simultaneously as the source in [14], but reaches a comparable brilliance of $4.1 \times 10^{19}$ ph s$^{-1}$ mm$^{-2}$ sr$^{-1}$ 0.1%BW$^{-1}$ at the longest center wavelength of 11.8 μm, $5.8 \times 10^{19}$ ph s$^{-1}$ mm$^{-2}$ sr$^{-1}$ 0.1%BW$^{-1}$ for a center wavelength of 10.8 μm, and exceeds $10^{20}$ ph s$^{-1}$ mm$^{-2}$ sr$^{-1}$ 0.1%BW$^{-1}$ for all outputs below 10 μm.

The 100-MHz repetition rate of the OPO means that it produces an essentially continuous-wave output, distinguishing it from kHz fiber sources [10], synchrotrons [12] and ns OPOs [15]. Such cw-modelocked performance allows the data acquisition rate of pump-probe experiments to be extremely high, enabling MHz-frequency lock-in detection which eliminates the technical noise typically present at kHz frequencies. The high repetition rate also presents exciting opportunities for rapid imaging in the fingerprint region, either in a micro-spectroscopy format as previously reported using synchrotron radiation [27], or in an active illumination format using raster scanning [28] or full-field infrared cameras [29].

Controlling the lengths of the pump and OPO resonators [30,31] enables the OP-GaP OPO to be easily configured as a frequency comb, with the prospect of implementing dual-comb spectroscopy [32] in the fingerprint region for the first time. Such a system could monitor changes in molecular functional groups at tens of kHz and with extremely high resolution, making it possible to observe multispecies reaction kinetics in real time.

By extending the reach of 1-μm synchronously pumped OPOs—previously limited by the transparencies of oxide nonlinear media such as LiNbO$_3$, LiB$_3$O$_5$ and KTiOPO$_4$—the OP-GaP source reported here represents a step-change in the practicality of molecular fingerprint-region broadband coherent sources, providing superior power, efficiency and repetition rate over competing approaches. We therefore expect OP-GaP technology to accelerate the development of new coherent spectroscopic applications of the mid-infrared in environmental sensing, security, healthcare, drug-quality screening, material science and optical histopathology.